# Minimizing Energy Consumption for Cooperative Network and Diversity Coded Sensor Networks

Gabriel E. Arrobo and Richard D. Gitlin
Department of Electrical Engineering
University of South Florida
Tampa, FL 33620
garrobo@mail.usf.edu, richgitlin@usf.edu

***Abstract*—In this paper, we present an approach to minimize the energy consumption of multihop wireless packet networks, while achieving the required level of reliability. We consider networks that use Cooperative Network Coding (CNC), which is a synergistic combination of Cooperative Communications and Network Coding. Our approach is to optimize and balance the use of forward error control, error detection, and retransmissions at the packet level for these networks. Additionally, we introduce Cooperative Diversity Coding (CDC), which is a novel means to code the information packets, with the aim of minimizing the energy consumed for coding operations. The performance of CDC is similar to CNC in terms of the probability of successful reception at the destination and expected number of correctly received information packets at the destination. However, CDC requires less energy at the source node because of its implementation simplicity. Achieving minimal energy consumption, with the required level of reliability is critical for the optimum functioning of many wireless sensor and body area networks. For representative applications, the optimized CDC or CNC network achieves ≥ 25% energy savings compared to the baseline CNC scheme.**

## I. INTRODUCTION

Achieving minimal energy consumption, with the required level of reliability is critical for the proper functioning of many wireless sensor and body area networks. In this paper we will address this challenge for advanced network architectures including *Cooperative Network Coding* (CNC) [1], which synergistically combines *Cooperative Communications* and *Network Coding,* to improve network performance by providing high throughput and overcoming packet losses. *Cooperative Communications* (CC) [2] improves the reliability of wireless links because the receiver processes data from multiple relays and by properly combining this data, the receiver can make more reliable decisions about the transmitted information. *Network Coding* (NC) [3] increases the throughput of networks by combining received packets, at intermediate nodes, from multiple links. It has been shown that NC also improves throughput in "noisy," or lossy, networks [1], [4]–[7]. However, in all of these network architectures, coded (parity) packets are transmitted to overcome wireless channel impairments. This increases network reliability at the expense of increasing the transmitted energy. We will address the design tradeoffs in optimizing the use of error control and retransmissions to optimize performance.

For CNC systems, as long as, the destination receives a sufficient number of error-free, innovative (linearly independent) combination packets, the original (source) packets may be properly recovered at the destination. There are two ways to implement NC; the first is through a centralized scheme, where the coding coefficients are assigned to the nodes by a central node. Complete linearly independency of the coded packets can be achieved with this methodology; however, the network topology needs to be known by all the nodes. The second way, which is the most commonly used, is through a decentralized scheme where each node randomly chooses the coding coefficients. This scheme is known as Random Linear Network Coding (RLNC) [8].

The paper is organized as follows. In section II, we summarize the prior work on cooperative network coding and retransmissions for wireless sensor networks. Our approach to improving the performance of wireless sensor networks using cooperative network coding, while minimizing energy consumption is presented in Section III. Section IV presents simulation results of the effect of cooperative network coding with retransmission(s) in wireless sensor networks. Finally, in Section V we present our conclusions.

## II. LITERATURE REVIEW

### *A. Cooperative Network Coding (CNC)*

Cooperative Network Coding (see Fig. 1 and Table I) was introduced in [1] and combines cluster-based cooperative communication and network coding with the aim of improving the communication reliability. The authors study the performance of this scheme in terms of probability of successful reception at the destination and the expected number of correctly received information packets at the destination. Through a mathematical analysis, the authors concluded that the number of nodes per cluster should be 15, when the number of original packets is 10. That is a network coding rate of 2/3. Also, they found that the optimal connectivity of the nodes should be 8 and that the throughput of this scheme is invariant with the number of hops (clusters). The authors compared this scheme with other three schemes: 1) no-cooperation and no-network coding; 2) cooperation and no-network coding; and 3) no-cooperation and network coding.

TABLE I. COOPERATIVE NETWORK CODING PARAMETERS [1]

| Item | Description |
|---|---|
| $n_i$ | Number of nodes in cluster *i* |
| $k$ | Number of clusters between the source and destination nodes |
| $r_{ij}$ | Number of nodes in cluster *i*+1 that are connected to node (*i,j*) |
| $r_s$ | Number of nodes in cluster 1 that are connected to the source node |
| $m$ | Number of information packets |
| $m'$ | Number of combination (coded) packets |
| $P_{(x)(y)}$ | Probability of link transmission loss between nodes *x* and *y* |

The effect of link-level feedback and retransmissions on the performance of Cooperative Network Coding was analyzed in [9]. The authors found that by having a retransmission in the last cluster (cluster *k*), the performance of Cooperative Network Coding can be improved when the number of nodes per cluster is low ($m \approx n_i$ for $\forall$ *i*). The analysis considers that all the nodes in the last cluster make a retransmission.

With CNC, before the source transmits the information to the destination, the nodes that are in the source-destination route/path recruit other nodes to form clusters [1], [10]. Then, the source node creates *m'* combination packets from a block of *m* original packets, where $m' \geq m$. The original packets are combined via operations in a Galois Field $GF(2^q)$, whose elements are $\{1, 2, 3, \ldots, 2^q\}$, as follows:

$$y_{S_j} = \sum_{l=1}^{m} c_{jl} x_l \,, \qquad j \in \{1, 2, 3, \ldots, m'\} \tag{1}$$

where $y_{sj}$ and $x_l$ are the combination (coded) packets and original packets, respectively, and the coefficients $c_{jl}$ are randomly chosen from $GF(2^q)$. The $GF(2^q)$ elements are $\{0, 1, 2, \ldots, 2^q - 1\}$ and each element (also known as symbol) is the representation of *q* bits. The $c_{jl}$ coefficients are embedded in the packet's header.

The energy required to network code a packet ($y_{sj}$) is [11]:

$$E_{NC} = mE_{LFSR} + \frac{L}{q}(mE_{MUL} + (m-1)E_{ADD}) \tag{2}$$

where $E_{LFSR}$ is the energy required to generate the random coefficients using a linear feedback shift register (LFSR), *L* is the packet length in bits, $E_{MUL}$ is the energy required to multiply a random coefficient and the packet (portion of the packet that depends on the Galois filed size) and $E_{ADD}$ is the energy required to add the results of two multiplication processes. Since with Network Coding, all the packets are coded, the energy required for each node to code *m'* packets is:

$$E_{NODE_{NC}} = m'E_{NC}$$

$$E_{NODE_{NC}} = m'\left(mE_{LFSR} + \frac{L}{q}(mE_{MUL} + (m-1)E_{ADD})\right) \tag{3}$$

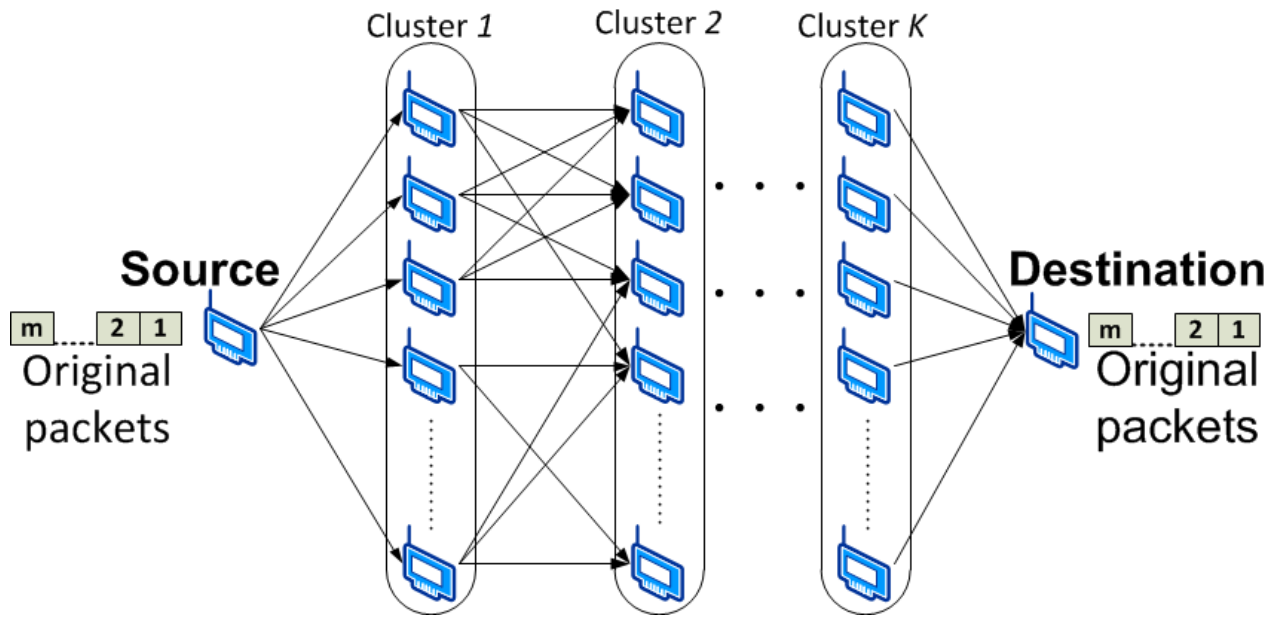

Figure 1. Cooperative Network Coding model.

The linear independency of the coded packets is a function of the field size. Thus, the expected number of transmitted packets until *m* linearly independent combination packets, are generated using RLNC, can be calculated as [11]:

$$M' = \sum_{i=1}^{m} \frac{1}{1 - \left(\frac{1}{2^q}\right)^i} \tag{4}$$

A typical field size is $GF(2^8)$. From (4), we can calculate the average probability $p_l$ of the *m'* combination packets being linearly independent:

$$p_l = \frac{m}{M'} \tag{5}$$

In Table II, we present the minimum number of transmitted combination packets, *m'*, needed to have *m* linearly independent packets for a field size equal to 8. Also, we calculate, $p_l$, the probability of linear independency of the transmitted combination packets. As we can see with RLNC, the source node needs to transmit a number of combination packets *m'* that is at least the smallest integer not less than *M'*.

$$m' = \lceil M' \rceil = \left\lceil \sum_{i=1}^{m} \frac{1}{1 - \left(\frac{1}{2^q}\right)^i} \right\rceil \tag{6}$$

TABLE II. MINIMUM NUMBER OF TRANSMITTED PACKETS (*M'*) AND PROBABILITY OF LINEAR INDEPENDENCY OF THE TRANSMITTED PACKETS

| Metric | Minimum number of transmitted packets and probability of their linear independency | | | |
|---|---|---|---|---|
| | *m=2* | *m=5* | *m=10* | *m=20* |
| $m'$ | 3 | 6 | 11 | 21 |
| $p_l$ | 99.8035% | 99.9213% | 99.9606% | 99.9803% |

Depending upon the degree of connectivity between the nodes in the first cluster and the source node, each node in cluster 1 (see Fig. 1) can correctly receive, on average, up to $m'{*}r_s/m$ combination packets if there are no transmission losses or errors, where $r_s$ is the number of nodes in cluster one that are connected to the source node (we assume that the $r_s$ nodes connected to the source node are uniformly distributed for each transmission of a combination packet). However, because of the channel characteristics (probability of transmission loss), some of combination packets may not be received. By combining the received packets, each node in cluster 1 creates a new combination packet and transmits it to the next cluster. In general, node *j* in the cluster *i* creates and transmits to nodes in cluster $i$+1 a combination packet from the received combination packets as follows:

$$y_{ij} = \sum_{l=1}^{m_j} c_{ijl} y_{i-1,l} \qquad j \in \{1,2,3,\dots,n_i\} \tag{7}$$

where $y_{ij}$ and $y_{i\text{-}1,l}$ are the transmitted combination packets and received combination packets, respectively, $m_j$ is the number of combination packets received by node $j$ in cluster $i$ from nodes in cluster $i-1$ and the coefficients $c_{ijl}$ are randomly chosen from $GF(2^q)$.

At the destination, the destination node needs to receive at least $m$ linearly independent combination packets from nodes in cluster $K$ to be able to recover the original information. Decoding could be done by block decoding or Gaussian elimination [6] applied to the matrix formed by the packets header to determine the original packets $\{x_l\}$.

$$\begin{pmatrix} c_{K11} & c_{K12} & \cdots & c_{K1m} \\ c_{K21} & c_{K22} & \cdots & c_{K2m} \\ \vdots & \vdots & \ddots & \vdots \\ c_{Km1} & c_{Km2} & \cdots & c_{Kmm} \end{pmatrix} \begin{pmatrix} x_1 \\ x_2 \\ \vdots \\ x_m \end{pmatrix} = \begin{pmatrix} y_1 \\ y_2 \\ \vdots \\ y_m \end{pmatrix} \tag{8}$$

Figure 1 shows the Cooperative Network Coding model that was described in the above paragraphs.

In order to achieve our goal of achieving minimal energy consumption, with the required level of reliability, we study the effect of the linear independency of the combination packets for multihop wireless networks and we propose a method to selectively retransmit combination packets from the last cluster where the combination packets have full rank. That is, since each node in a cluster transmits only one coded packet and all the nodes in a cluster cooperate to transmit linear independent coded packets, full rank in a cluster is achieved when at least $n_i$ nodes in cluster $i$ transmit $m$ linear independent packets, where $n_i \geq m$. By using this selectivity of the retransmitted packets, we can minimize the average energy consumed by each node and the energy required by the source node to create and transmit combination packets.

Reference [9] extended the work done in [1] and analyzed the effect of link-level feedback and retransmission from the last cluster on the performance of cooperative network coding. As in these references, in this paper we consider the probability of success and expected number of correctly received packets at the destination, along with energy consumption, as the performance metrics for our statistical analysis.

Cooperative Network Coding provides high reliability, due to the diversity of routes/paths, which increases the probability that the destination receives a sufficient number of linearly independent packets to be able to decode the original information. Consequently, Cooperative Network Coding offers robust protection against failures of links and/or nodes. In case that the destination receives less than the minimum number of linearly independent coded packets, selective retransmission from the last cluster that has full rank (at least $m$ linearly independent combination packets) can be made to avoid any retransmission from the source node. This feature of Cooperative Network Coding is very useful for multihop networks.

### B. *Related Work*

To overcome the effect of lost, or errored, packets, communication systems use retransmissions to increase the probability of successful delivery of a message. The retransmissions can be done end-to-end or link-by-link. In end-to-end retransmissions, the destination node acknowledges (ACK) the reception of a packet and if the transmitted packet or the ACK is lost, the source node retransmits the packet. This is used in the transport layer (e.g. TCP protocol). In link-by-link retransmissions, a transmitted packet is acknowledged on a link basis. That is, the source transmits a packet to the next node that is in the path towards the destination. Then, the node acknowledges the packet to the source. If the packet is lost, the source retransmits the packet. If not, the node sends the packet to the following node in the path to the destination and the following node acknowledges successful reception of the packet. If the packet is lost, the packet is retransmitted. This process continues until the penultimate node in the path sends the packet to the destination node and waits for an acknowledgement. If the packet is lost, the penultimate node retransmits the packet. Link-by-link retransmission is implemented at the link layer through Automatic Repeat reQuest (ARQ).

Link-by-link retransmission provides higher reliability compared to the end-to-end retransmission. However, the main drawback of this approach is its latency because of retransmissions and also requires buffers and timers. As a result the load on the network is very high, especially with bad channel conditions, and the capacity is reduced because of the need of reverse channel.

Considering the advantages and disadvantages of end-to-end and link-by-link retransmissions, a mixed approach is presented in this paper with the aim of minimizing the consumed energy for coding and transmitting packets while optimizing the performance of multihop wireless networks that use Cooperative Network Coding.

In this paper, by a statistical study, we minimize the number of transmitted packets by each cluster and the source node while maximizing the probability of successful reception at the destination and minimizing the number of retransmitted packets and the energy consumption of the nodes.

## III. THE EFFECT OF RETRANSMISSIONS IN COOPERATIVE NETWORK AND DIVERSITY CODING

In the previous sections, we present Cooperative Network Coding where the source and the nodes in a cluster randomly choose the coding coefficients. However, with the aim of minimizing the energy consumed by the source due to coding operations, we also study the performance of cooperative diversity coding (CDC). Cooperative Diversity Coding operates similarly as to Cooperative Network Coding but the difference is in the method by which the source chooses the coding coefficients. The source uses Diversity Coding [12], an efficient technique to code packets. For CNC and CDC, the source has two alternatives to create the combination packets. Either create packets by randomly choosing the coding coefficients or by taking known coefficients from the Vandermonde matrix. By randomly choosing the coefficients, linearly independency of the combination packets is not guaranteed as we can see in Table II. Moreover, the linearly independence of the combination packets depends on the Galois field size. The higher the field size the higher is the probability of linear independence of the combination packets, we can verify this using (4). On the other hand, by selecting known coefficients from a Vandermonde matrix, it is guaranteed that all combination packets will be linearly independent at the source. Moreover, since the coefficients are known, the computational complexity is reduced because the Vandermonde matrix coefficients are stored in the sensor memory. Also, no extra circuitry is needed (e.g. shift registers) as is the case for RLNC. The simplicity of using Vandermonde matrix coefficients was implemented in Diversity Coding [12], a forerunner of Network Coding. In diversity coding, the coding coefficients ($\beta_{ij}$) are calculated as follow:

$$\beta_{ij} = \alpha^{(i-1)(j-1)} \tag{9}$$

where α is a primitive element of a Galois field $GF(2^q)$, and the indexes $i \in \{1, 2, \ldots, m\}$ and $j \in \{1, 2, \ldots, m'\}$. We can present the coding coefficients in a matrix form as follows:

$$\beta_{ij} = \begin{pmatrix} 1 & 1 & 1 & \cdots & 1 \\ 1 & \alpha & \alpha^2 & \cdots & \alpha^{(m-1)} \\ 1 & \alpha^2 & \alpha^4 & \cdots & \alpha^{(m-1)2} \\ \vdots & \vdots & \vdots & \ddots & \vdots \\ 1 & \alpha^{(m'-1)} & \alpha^{2(m'-1)} & \cdots & \alpha^{(m-1)(m'-1)} \end{pmatrix} \tag{10}$$

Thus, the source uses the $\beta_{ij}$ coefficients to calculate the coded packets. Therefore, (1) becomes:

$$y_{S_j} = \sum_{l=1}^{m} \beta_{jl} x_l, \qquad j \in \{1, 2, 3, \ldots, m'\} \tag{11}$$

As we previously mentioned, the nodes in a cluster still uses RLNC to combine the received packets and create a new combination packet.

Note that with CDC the combination packets lose their linear independency at the clusters, because the nodes in a cluster still use random network coding to create the new combination packets. The main advantage of CDC over CNC is that the source saves computation energy by creating coded packets using known coding coefficients (nor random coefficients), which are stored in the node's memory. Also, with CDC, only the additional (protection) packets are coded and the original information is transmitted uncoded. That is, the energy required to code a packet using Diversity Coding is calculated as:

$$E_{DC} = \frac{L}{q}(mE_{MUL} + (m-1)mE_{ADD}) \tag{12}$$

where $L$ is the packet length in bits, $E_{MUL}$ is the energy require to multiply a random coefficient and the packet (portion of the packet that depends on the Galois filed size) and $E_{ADD}$ is the energy required to add the results of two multiplication processes. Since with Diversity Coding, only the protection packets are coded, the energy required for the source node to code $m'$ packets is:

$$E_{NODE_{DC}} = (m' - m)E_{DC}$$

$$E_{NODE_{DC}} = (m' - m)\frac{L}{q}(mE_{MUL} + (m-1)mE_{ADD}) \tag{13}$$

As we can see from equations (3) and (13), the source node requires less energy when using Diversity Coding to create coded packets ($Esource_{DC}$). That is given in (14):

$$E_{SOURCE_{DC}} = E_{SOURCE_{NC}} - m'mE_{LFSR} - m\frac{L}{q}(mE_{MUL} + (m-1)mE_{ADD}) \tag{14}$$

where the second term on the right hand side of (14) is the energy savings for using known coding coefficients and the third term on the right hand side of (14) is the energy savings achieved for coding only the protection packets.

We can express the total number of transmitted packets in the network with CDC or CNC as:

$$E_{TOTAL} = m' + \prod_{i=1}^{K} n_i \tag{15}$$

In the following section we present the simulation results and a statistical analysis of those simulations with the aim of minimizing the energy required to transmit a block of information by minimizing the number of transmitted packets.

## IV. Simulation Results

In this work, we propose a method to minimize the energy consumption for Cooperative Network Coding and Cooperative Diversity Coding systems. We have simulated the effect of different parameters, such as: number of coded packets, coding coefficients at the source by using random coefficients or Vandermonde matrix coefficients, linear independency of the packets at each cluster (rank) to minimize the average energy consumed by the network while optimizing the network's throughput.

The results presented in the tables below were obtained through simulations by running 1,000 experiments. An experiment is considered successful when the sink was able to decode the information from the source. Additionally, we assumed that the network consists of 20 clusters, each cluster has a number of nodes equal to the number of combination packets, and the source has a block of information of 10 packets. Also, we assumed that the probability of link transmission loss is the same for all the links (this assumption is unrealistic, but it simplifies the study). We performed the network coding operations over a Galois field $GF(2^8)$ with packets size of 100 bytes.

As shown in Fig. 2, the source needs to transmit at least $m$+1 combination packets; otherwise the source needs to retransmit with very high probability. This is because the links between the source and the nodes in the first cluster are error prone.

Tables III and IV show the linear independency of the packets at each cluster for CNC and CDC for different connectivity parameters and probability of link transmission loss of 0.10, given that the source node transmitted 11 combination packets. Based on the statistical analysis, we can see that there is no need for making a retransmission from the source node because clusters 4 and 11, respectively, have full rank and the retransmission can be made from those clusters.

Figure 3 along with Tables V and VI show the most general case where full rank, that is at least *m* linearly independent correct packets, is achieved at a sufficient number of nodes including the last cluster, and a selective retransmission has to be made by the nodes in the last cluster for the destination to be able to decode the source's information. Fig. 3 shows the expected number of information packets decoded at the destination as a function of the number of combination packets. As noted, the source node should transmit at least $m+1$ combination packets to avoid retransmissions. Table V presents the results for cooperative network coding given that the probability of link transmission loss is 0.05, the connectivity among the nodes is 6 and the number of combination packets transmitted by the source is 11. In the worst case (when only 7 coded packets have been received at the destination node) 3 nodes in the last cluster need to retransmit a combination packet. A similar situation is shown in Table VI, where only 2 nodes in the last cluster need to retransmit. Considering these two examples, we can see that only 11 combination packets should be transmitted by each cluster (one combination packet per node) plus one retransmission (3 and 2 combination packets, respectively) from the last cluster (cluster 20) for the destination to be able to reliably decode the *m* original packets.

Moreover, Tables III-VI show that the skewness, a measure of the asymmetry of a probability distribution, is negative, which means that most of the values of the probability distribution lie to the left of the mean. For example, in Tables V and VI, the mean at the destination node is 9.88, which means that all the 10 original packets were recovered in an average of 98.8% of the simulations. Also, Tables V and VI show that most of the values lie to the right of the 9.88. In other words, the 10 original packets are recovered most of the time. This characteristic is omitted in analytical analyses because the analytical analyses only consider averages values.

Comparing with [1], where 15 combination packets should be transmitted to achieve full throughput, our approach reduces by 26% the energy consumed by the network. That is, the source node transmits 15 coded packets to the nodes in the first cluster. Then, the nodes in a cluster (15 nodes per cluster), transmits one coded packet to the nodes in the next cluster, and so on. So, using (15), we can calculate the number of transmitted packets to achieve full throughput at the destination, which is 315 packets for a 21-hop network. Using our approach, we only need to transmit 233 packets (Table VI) to achieve full throughput at the destination node. Moreover, since fewer packets are being received by each node, the energy required to create a new coded packet is reduced. In addition, extra energy savings (at the source node) are achieved when CDC is used.

Figure 4 shows the performance of CNC and CDC vs. the number of nodes per cluster. As expected, the performance of these two approaches increases when the number of nodes per cluster increases because there are more nodes in each cluster transmitting combination packets. However, increasing the number of nodes per cluster is not a preferred option because of the extra energy that is spent by the entire network. A better option is to retransmit from the last cluster, where the system still has full rank (linear independency of the combination packets).

We determined the full rank of the coding coefficients (linear independency among the packets) is lost in the first hop about ~98.5% - 99.9% of the time, depending on the connectivity among the nodes, the number of nodes per cluster, and the probability of link transmission loss. However, for 10 original data packets, and independently of the number of hops and the connectivity among the nodes, the probability of successful reception at destination is essentially unity when 14 and 16 coded packets are transmitted for a probability of transmission loss of 0.05 and 0.10, respectively. That is, a coding overhead of 40% for a probability of link transmission loss of 0.05 and 60% coding overhead for a probability of link transmission loss 0.10 will achieve full throughput. However, with the aim of minimizing the energy consumed by the nodes in transmitting coded packets, the source need only transmit about 10% - 30% coded packets and utilize retransmission by the nodes in the last cluster that has full rank (100% linear independency among the packets) to minimize energy utilization. Our statistical analysis has shown that most of the retransmissions are only required at the last cluster. In this way we optimize the energy consumed by each node and minimize the energy consumed by the source node. Moreover, we minimize the delay introduce by the retransmissions because no retransmission is done from the source node.

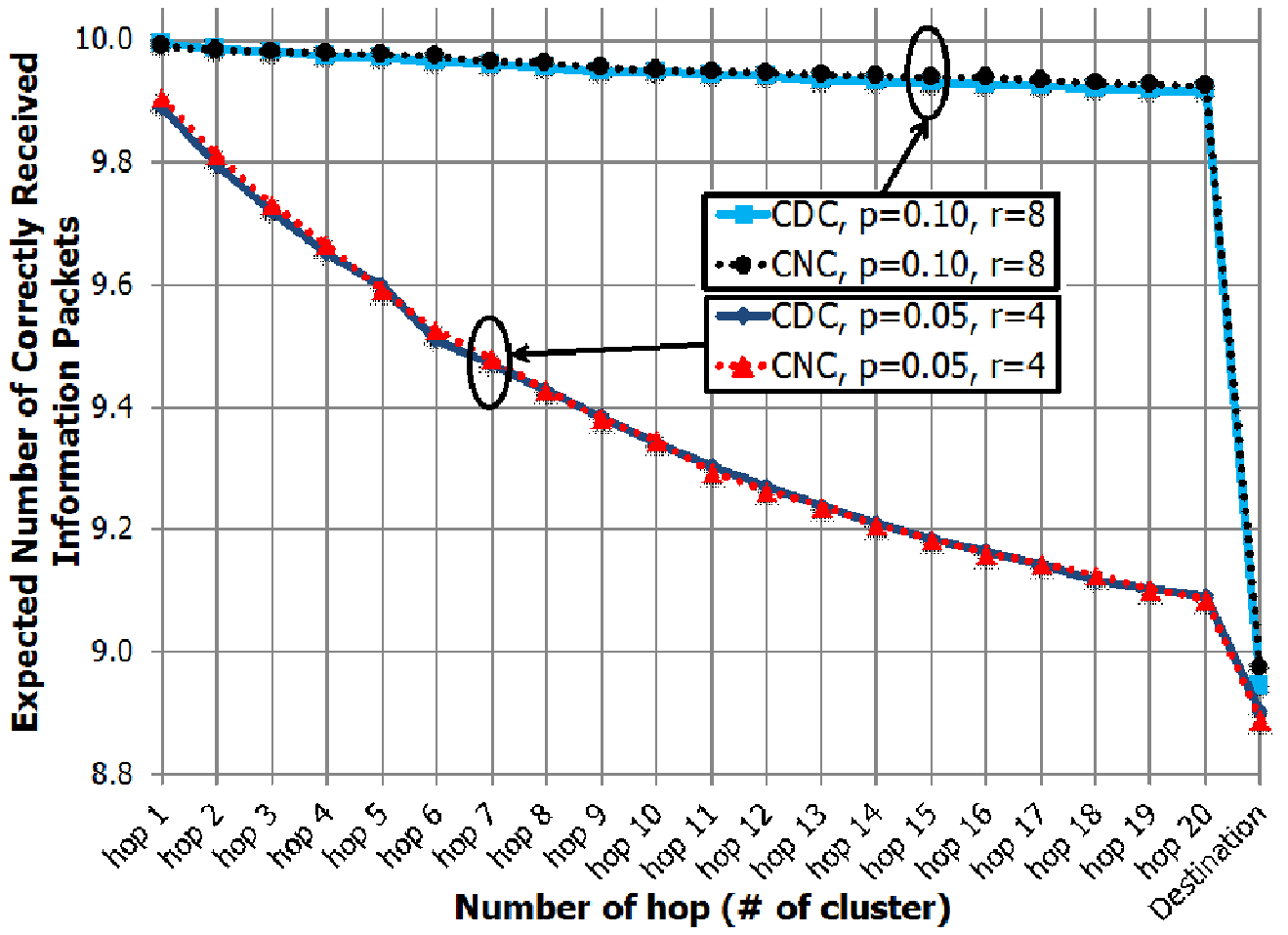

Figure 2. CNC and CDC performance for probability of link transmission loss equal to *p*, connectivity equal to *r* and $m=m'=10$ combination packets.

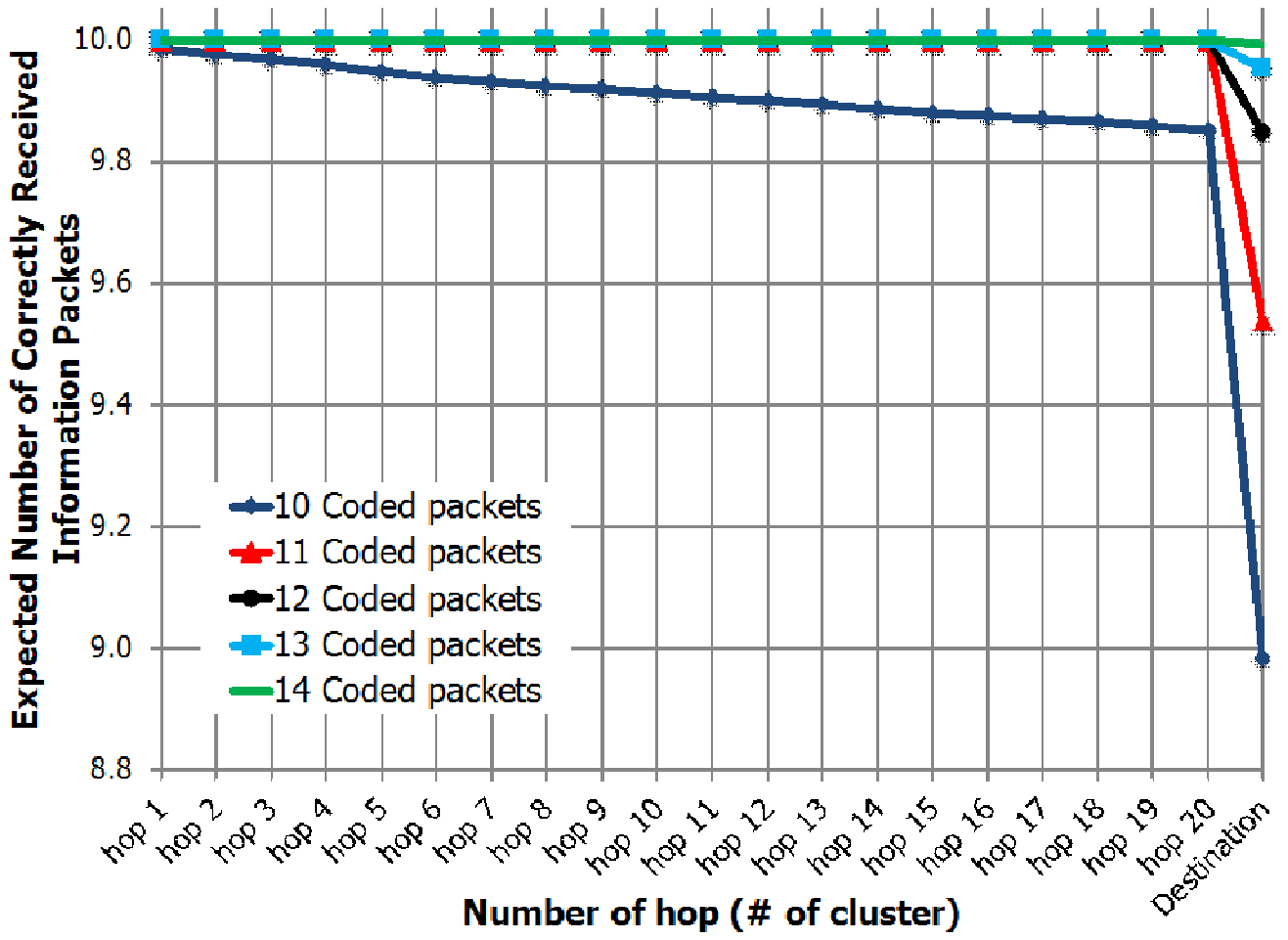

Figure 3. CNC performance for probability of link transmission loss equal to 0.10, connectivity equal to 6 and $m=10$ original packets.

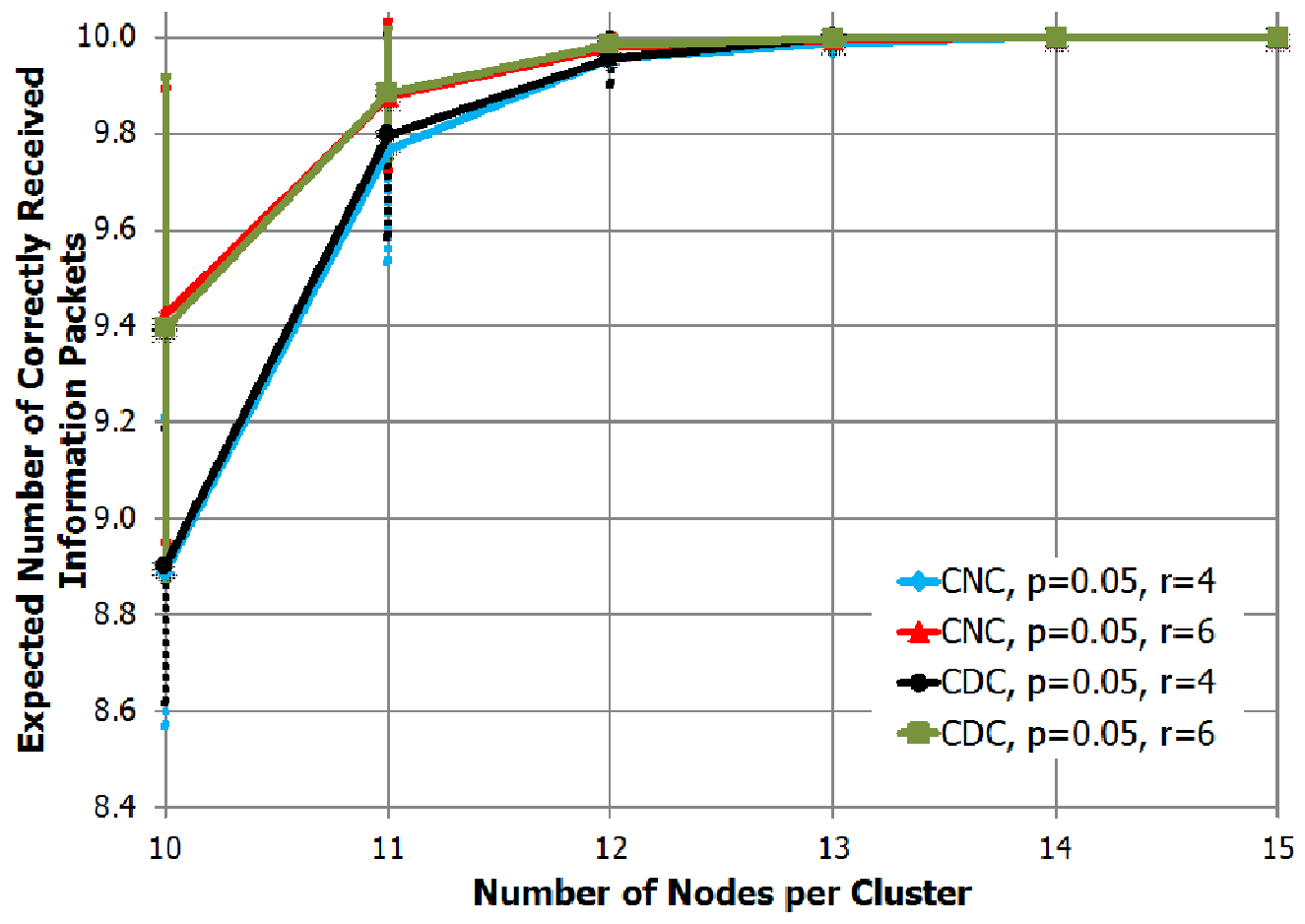

Figure 4. CNC and CDC performance vs. the number of nodes per cluster for probability of link transmission loss equal to 0.05, connectivity equal to r and *m*=10 original packets.

Since these are random events, minimum energy consumption can be achieved in practice by the nodes listening to the transmitted combination packets from their own cluster and calculating the rank of the system. Noted that the nodes that form a cluster are geographically close to each other and they can hear each other's packets with high probability.

In summary the above approach of selective retransmissions will minimize both the energy consumed by the network and the delay, while achieving the desired throughput.

## V. Conclusions

Our approach of selective retransmissions minimizes the energy consumed by multihop wireless packet networks that use Cooperative Network Coding (CNC) or a novel variant Cooperative Diversity Coding (CDC). By optimizing and balancing the use of forward error control, error detection, and retransmissions at packet level in such networks we can both minimize the energy consumption and network latency.

The energy savings obtained by our approaches (CNC and CDC) are about 26% compared to the baseline CNC approach. Further, our CDC approach further reduces the energy and complexity of the source node to create combination packets.

TABLE III. CNC PERFORMANCE FOR PROBABILITY OF LINK TRANSMISSION LOSS OF 0.10, CONNECTIVITY AMONG THE NODES EQUAL TO 6, AND 11 COMBINATION PACKETS

**Descriptive Statistics**

| | | hop 1 | hop 2 | hop 3 | hop 4 | hop 5 | hop 6 | hop 7 | hop 8 | hop 9 | hop 10 | hop 11 | hop 12 | hop 13 | hop 14 | hop 15 | hop 16 | hop 17 | hop 18 | hop 19 | hop 20 | Destination |
|---|---|---|---|---|---|---|---|---|---|---|---|---|---|---|---|---|---|---|---|---|---|---|
| N | Statistic | 1000 | 1000 | 1000 | 1000 | 1000 | 1000 | 1000 | 1000 | 1000 | 1000 | 1000 | 1000 | 1000 | 1000 | 1000 | 1000 | 1000 | 1000 | 1000 | 1000 | 1000 |
| Range | Statistic | 0 | 0 | 0 | 0 | 1 | 1 | 1 | 1 | 1 | 1 | 1 | 1 | 1 | 1 | 1 | 1 | 1 | 1 | 1 | 1 | 5 |
| Minimum | Statistic | 10 | 10 | 10 | 10 | 9 | 9 | 9 | 9 | 9 | 9 | 9 | 9 | 9 | 9 | 9 | 9 | 9 | 9 | 9 | 9 | 5 |
| Maximum | Statistic | 10 | 10 | 10 | 10 | 10 | 10 | 10 | 10 | 10 | 10 | 10 | 10 | 10 | 10 | 10 | 10 | 10 | 10 | 10 | 10 | 10 |
| Mean | Statistic | 10.00 | 10.00 | 10.00 | 10.00 | 10.00 | 10.00 | 10.00 | 10.00 | 10.00 | 10.00 | 10.00 | 10.00 | 10.00 | 10.00 | 10.00 | 10.00 | 10.00 | 10.00 | 10.00 | 10.00 | 9.54 |
| | Std. Error | .000 | .000 | .000 | .000 | .001 | .001 | .001 | .001 | .001 | .001 | .001 | .001 | .001 | .001 | .001 | .001 | .001 | .001 | .001 | .001 | .025 |
| Std. Deviation | Statistic | .000 | .000 | .000 | .000 | .032 | .032 | .032 | .032 | .032 | .032 | .032 | .032 | .032 | .032 | .032 | .032 | .032 | .032 | .032 | .032 | .798 |
| Variance | Statistic | .000 | .000 | .000 | .000 | .001 | .001 | .001 | .001 | .001 | .001 | .001 | .001 | .001 | .001 | .001 | .001 | .001 | .001 | .001 | .001 | .637 |
| Skewness | Statistic | . | . | . | . | -31.623 | -31.623 | -31.623 | -31.623 | -31.623 | -31.623 | -31.623 | -31.623 | -31.623 | -31.623 | -31.623 | -31.623 | -31.623 | -31.623 | -31.623 | -31.623 | -1.884 |
| | Std. Error | . | . | . | . | .077 | .077 | .077 | .077 | .077 | .077 | .077 | .077 | .077 | .077 | .077 | .077 | .077 | .077 | .077 | .077 | .077 |

TABLE IV. CDC PERFORMANCE FOR PROBABILITY OF LINK TRANSMISSION LOSS OF 0.10, CONNECTIVITY AMONG THE NODES EQUAL TO 6, AND 11 COMBINATION PACKETS

**Descriptive Statistics**

| | | hop 1 | hop 2 | hop 3 | hop 4 | hop 5 | hop 6 | hop 7 | hop 8 | hop 9 | hop 10 | hop 11 | hop 12 | hop 13 | hop 14 | hop 15 | hop 16 | hop 17 | hop 18 | hop 19 | hop 20 | Destination |
|---|---|---|---|---|---|---|---|---|---|---|---|---|---|---|---|---|---|---|---|---|---|---|
| N | Statistic | 1000 | 1000 | 1000 | 1000 | 1000 | 1000 | 1000 | 1000 | 1000 | 1000 | 1000 | 1000 | 1000 | 1000 | 1000 | 1000 | 1000 | 1000 | 1000 | 1000 | 1000 |
| Range | Statistic | 0 | 0 | 0 | 0 | 0 | 0 | 0 | 0 | 0 | 0 | 0 | 1 | 1 | 1 | 1 | 1 | 1 | 1 | 1 | 1 | 3 |
| Minimum | Statistic | 10 | 10 | 10 | 10 | 10 | 10 | 10 | 10 | 10 | 10 | 10 | 9 | 9 | 9 | 9 | 9 | 9 | 9 | 9 | 9 | 7 |
| Maximum | Statistic | 10 | 10 | 10 | 10 | 10 | 10 | 10 | 10 | 10 | 10 | 10 | 10 | 10 | 10 | 10 | 10 | 10 | 10 | 10 | 10 | 10 |
| Mean | Statistic | 10.00 | 10.00 | 10.00 | 10.00 | 10.00 | 10.00 | 10.00 | 10.00 | 10.00 | 10.00 | 10.00 | 10.00 | 10.00 | 10.00 | 10.00 | 10.00 | 10.00 | 10.00 | 10.00 | 10.00 | 9.60 |
| | Std. Error | .000 | .000 | .000 | .000 | .000 | .000 | .000 | .000 | .000 | .000 | .000 | .001 | .001 | .001 | .001 | .001 | .001 | .001 | .001 | .001 | .022 |
| Std. Deviation | Statistic | .000 | .000 | .000 | .000 | .000 | .000 | .000 | .000 | .000 | .000 | .000 | .032 | .032 | .032 | .032 | .032 | .032 | .032 | .032 | .032 | .696 |
| Variance | Statistic | .000 | .000 | .000 | .000 | .000 | .000 | .000 | .000 | .000 | .000 | .000 | .001 | .001 | .001 | .001 | .001 | .001 | .001 | .001 | .001 | .485 |
| Skewness | Statistic | . | . | . | . | . | . | . | . | . | . | . | -31.623 | -31.623 | -31.623 | -31.623 | -31.623 | -31.623 | -31.623 | -31.623 | -31.623 | -1.730 |
| | Std. Error | . | . | . | . | . | . | . | . | . | . | . | .077 | .077 | .077 | .077 | .077 | .077 | .077 | .077 | .077 | .077 |

TABLE V. CNC PERFORMANCE FOR PROBABILITY OF LINK TRANSMISSION LOSS OF 0.05, CONNECTIVITY AMONG THE NODES EQUAL TO 6, AND 11 COMBINATION PACKETS

**Descriptive Statistics**

| | | hop 1 | hop 2 | hop 3 | hop 4 | hop 5 | hop 6 | hop 7 | hop 8 | hop 9 | hop 10 | hop 11 | hop 12 | hop 13 | hop 14 | hop 15 | hop 16 | hop 17 | hop 18 | hop 19 | hop 20 | Destination |
|---|---|---|---|---|---|---|---|---|---|---|---|---|---|---|---|---|---|---|---|---|---|---|
| N | Statistic | 1000 | 1000 | 1000 | 1000 | 1000 | 1000 | 1000 | 1000 | 1000 | 1000 | 1000 | 1000 | 1000 | 1000 | 1000 | 1000 | 1000 | 1000 | 1000 | 1000 | 1000 |
| Range | Statistic | 0 | 0 | 0 | 0 | 0 | 0 | 0 | 0 | 0 | 0 | 0 | 0 | 0 | 0 | 0 | 0 | 0 | 0 | 0 | 0 | 3 |
| Minimum | Statistic | 10 | 10 | 10 | 10 | 10 | 10 | 10 | 10 | 10 | 10 | 10 | 10 | 10 | 10 | 10 | 10 | 10 | 10 | 10 | 10 | 7 |
| Maximum | Statistic | 10 | 10 | 10 | 10 | 10 | 10 | 10 | 10 | 10 | 10 | 10 | 10 | 10 | 10 | 10 | 10 | 10 | 10 | 10 | 10 | 10 |
| Mean | Statistic | 10.00 | 10.00 | 10.00 | 10.00 | 10.00 | 10.00 | 10.00 | 10.00 | 10.00 | 10.00 | 10.00 | 10.00 | 10.00 | 10.00 | 10.00 | 10.00 | 10.00 | 10.00 | 10.00 | 10.00 | 9.88 |
| | Std. Error | .000 | .000 | .000 | .000 | .000 | .000 | .000 | .000 | .000 | .000 | .000 | .000 | .000 | .000 | .000 | .000 | .000 | .000 | .000 | .000 | .012 |
| Std. Deviation | Statistic | .000 | .000 | .000 | .000 | .000 | .000 | .000 | .000 | .000 | .000 | .000 | .000 | .000 | .000 | .000 | .000 | .000 | .000 | .000 | .000 | .394 |
| Variance | Statistic | .000 | .000 | .000 | .000 | .000 | .000 | .000 | .000 | .000 | .000 | .000 | .000 | .000 | .000 | .000 | .000 | .000 | .000 | .000 | .000 | .155 |
| Skewness | Statistic | . | . | . | . | . | . | . | . | . | . | . | . | . | . | . | . | . | . | . | . | -3.794 |
| | Std. Error | . | . | . | . | . | . | . | . | . | . | . | . | . | . | . | . | . | . | . | . | .077 |

TABLE VI. CDC PERFORMANCE FOR PROBABILITY OF LINK TRANSMISSION LOSS OF 0.05, CONNECTIVITY AMONG THE NODES EQUAL TO 6, AND 11 COMBINATION PACKETS

**Descriptive Statistics**

| | | hop 1 | hop 2 | hop 3 | hop 4 | hop 5 | hop 6 | hop 7 | hop 8 | hop 9 | hop 10 | hop 11 | hop 12 | hop 13 | hop 14 | hop 15 | hop 16 | hop 17 | hop 18 | hop 19 | hop 20 | Destination |
|---|---|---|---|---|---|---|---|---|---|---|---|---|---|---|---|---|---|---|---|---|---|---|
| N | Statistic | 1000 | 1000 | 1000 | 1000 | 1000 | 1000 | 1000 | 1000 | 1000 | 1000 | 1000 | 1000 | 1000 | 1000 | 1000 | 1000 | 1000 | 1000 | 1000 | 1000 | 1000 |
| Range | Statistic | 0 | 0 | 0 | 0 | 0 | 0 | 0 | 0 | 0 | 0 | 0 | 0 | 0 | 0 | 0 | 0 | 0 | 0 | 0 | 0 | 2 |
| Minimum | Statistic | 10 | 10 | 10 | 10 | 10 | 10 | 10 | 10 | 10 | 10 | 10 | 10 | 10 | 10 | 10 | 10 | 10 | 10 | 10 | 10 | 8 |
| Maximum | Statistic | 10 | 10 | 10 | 10 | 10 | 10 | 10 | 10 | 10 | 10 | 10 | 10 | 10 | 10 | 10 | 10 | 10 | 10 | 10 | 10 | 10 |
| Mean | Statistic | 10.00 | 10.00 | 10.00 | 10.00 | 10.00 | 10.00 | 10.00 | 10.00 | 10.00 | 10.00 | 10.00 | 10.00 | 10.00 | 10.00 | 10.00 | 10.00 | 10.00 | 10.00 | 10.00 | 10.00 | 9.88 |
| | Std. Error | .000 | .000 | .000 | .000 | .000 | .000 | .000 | .000 | .000 | .000 | .000 | .000 | .000 | .000 | .000 | .000 | .000 | .000 | .000 | .000 | .012 |
| Std. Deviation | Statistic | .000 | .000 | .000 | .000 | .000 | .000 | .000 | .000 | .000 | .000 | .000 | .000 | .000 | .000 | .000 | .000 | .000 | .000 | .000 | .000 | .368 |
| Variance | Statistic | .000 | .000 | .000 | .000 | .000 | .000 | .000 | .000 | .000 | .000 | .000 | .000 | .000 | .000 | .000 | .000 | .000 | .000 | .000 | .000 | .135 |
| Skewness | Statistic | . | . | . | . | . | . | . | . | . | . | . | . | . | . | . | . | . | . | . | . | -3.298 |
| | Std. Error | . | . | . | . | . | . | . | . | . | . | . | . | . | . | . | . | . | . | . | . | .077 |